\documentstyle[]{mn}
\input psfig.sty
\input amssymb.sty

\title[Temperature profile in clusters]
{Steady-state conduction-driven temperature profile in clusters of galaxies}

\author[S. Dos Santos]
{S. Dos Santos,$^{1,2}$ \\
$^1$ X-ray Astronomy Group, Department of Physics and  
Astronomy, University of Leicester, Leicester LE1 7RH, UK \\
$^2$ Institut d'Astrophysique de Paris, CNRS, Universit\'e Pierre et
Marie Curie, 98bis Bd Arago, F-75014 Paris, France\\ 
$^3$ e-mail: ssa@star.le.ac.uk \\}

\date{Accepted ???. Received ????; in original form ????}
\pagerange{\pageref{firstpage}--\pageref{lastpage}}
\pubyear{2000}

\begin{document}

\maketitle
\label{firstpage}

\begin{abstract}

The temperature profile (TP) of the intracluster medium (ICM) is of
primeval importance for deriving the dynamical 
parameters of the largest equilibrium systems known in the universe,
in particular their total mass profile. Analytical models of the ICM
often assume that the ICM is isothermal or parametrize the TP with a
polytropic index $\gamma_p$. This parameter is ajusted to observations,
but has in fact poor physical meaning for  values other than
1 or 5/3, when considering monoatomic gases. 
In this article, I present a theoretical model of a relaxed cluster
where the  TP is instead structured by electronic thermal
conduction. Neglecting cooling and heating terms, the stationnary energy 
conservation equation reduces to a second order differential equation,
 whose resolution
requires two boundary conditions, taken here as the inner radius and
the ratio between inner and outer temperature. Once these two
constants are chosen, the TP has a fixed analytical expression, 
which reproduces nicely the observed ``universal''
TP obtained by Markevitch et al. (1998) from ASCA data. 
Using observed X-ray surface brightnesses for two hot clusters with
spatially resolved TP, the local polytropic index and the hot gas
fraction profile are predicted and
compare very well with ASCA observations (Markevitch et al
1999). Moreover, the total density 
profile derived from observed X-ray surface brightness, hydrostatic
equilibrium and the conduction-driven TP is very well fit by three
analytical 
profiles found to describe the structure of galactic or cluster
halos in numerical simulations of collisionless matter 
(Hernquist, 1991; Navarro et al. 1995, 1997; Burkert 1995). 

The suppression of the heat conduction several orders of magnitude
below the Spitzer rate is an important assumption of the cooling-flow
models, in order to ensure the thermal instability ability to trigger
further condensation and cooling of density perturbations, although no
definitive theoretical picture of this reduction has yet been put
forward. In consequence, electronic  heat conduction has seldom been
considered for the structure of the main volume of the cluster,
outside the cooling flow radius. However, the physical situation
outside the cooling flow differs widely from the one inside, the
temperature gradient being much shallower, the magnetic field
intensity much smaller (as shown by the Faraday rotation measures and
predicted by Soker \& Sarazin, 1990) and the cooling time higher than
the mean age of the 
structure. Thus, it is not obvious that the mechanism
reducing the heat flux in the cooling flow is as highly effective in
the main body of a cluster. If the TP decline in clusters is confirmed
by the new 
generation of X-ray telescopes (Chandra and XMM-Newton), this simple
conduction-driven model of the cluster ICM equilibrium could give
useful insights on the physical situation in this region and the
predicted shape of the TP (related to the temperature dependance of
the heat flux for a collisionally-ionised plasma) will be tested
directly against observations.

\end{abstract}

\begin{keywords} 
hydrodynamics -- conduction -- magnetic fields -- methods: analytical
-- galaxies: clusters: general -- dark matter -- X-rays: general 
\end{keywords}

\section{Introduction}
\label{intro}

The numerical simulations of the formation of clusters in a realistic
cosmological frame seem to have reached a fair state, since the first runs of
two-fluid P3MSPH simulations by Evrard (1988, 1990).
Recently, Frenk et al. (1999) have compared, using 12
different codes, the final output of the simulation of an X-ray cluster in a
CDM universe, and have found that the overall agreement is impressive, except
for quantities requiring an enhanced resolution, such as the total X-ray 
luminosity. Moreover, although agreement does not guarantee correctness (as
scaling law models can only give the evolution of an ideal population of
standard clusters), the results of simulations are remarkably fitted by
predictions of approximate analytic models (Navarro, Frenk \& White 1995
hereafter NFW95, Eke, Navarro \& Frenk 1998, Bryan \& Norman 1998). Thus, it
seems that numerical simulations of clusters are a reliable implementation of
the physical processes invoked for the adiabatic formation of clusters. 

However, these simulations make the simplifying assumption that only gravity,
pressure gradients and hydrodynamical shocks are important in the evolution
of clusters. Moreover, when the results of observed statistical properties of
X-ray clusters are compared to simulations, one of the most basic observed
relations, namely the $L_X-T$ relation, cannot be reproduced, being shallower
than the observations (see Bryan \& Norman, 1998).  This relation is of
fundamental importance since it links the mean temperature (thought as
imposed by the total mass) and the luminosity (which goes as the baryonic
density squared times the cube of a characteristic radius of the X-ray
halo). Thus, the $L_X-T$ relation shows the changes with temperature in
the equilibrium state of the baryonic gas in the underlying dark matter
potential. The same flattening of the $L_X-T$ relation, which corresponds to
a steepening of the luminosity function (as compared to observations) was
found by Kaiser (1991) and Evrard \& Henry (1991) in the validation of
adiabatic scaling law models. Both found that an early preheating phase
(maybe galactic feedback or quasar formation), enhancing the initial adiabat
of the gas before cluster formation, could solve this problem, decreasing the
final density and thus the final luminosity. Since low virial temperature
systems are more sensitive to this phenomenon, the $L_X-T$ relation is
steepened in the correct way. Such a
scenario has been crudely incorporated in numerical simulations by assuming a
higher initial temperature (Evrard 1990, NFW95) and effectively produces less
dense clusters with larger X-ray core radii. Recently, such core
X-ray properties of groups of 
galaxies as compared to clusters, have been interpreted as the
presence of a  minimum
entropy threshold, higher than the only gravitational processes could
have produced (Ponman, Cannon \& Navarro
1998). Finally, semi-analytical
models of structure formation, applied to groups and clusters of
galaxies, have shown that an early injection by supernovae or quasars
can reproduce the self-similarity breaking for structures with virial
temperature smaller than $\sim 3 \, \rm keV$ (see {\it e.g.}
Cavaliere, Menci \& Tozzi 1997,1998; Wu, Fabian \& Nulsen 1999;
Valageas \& Silk 1999; Bower et 
al. 2000). Whatever the details of the  equilibrium of the gas in the
gravitational potential, this energy excess seems to be of the order
of $\sim 1 \, \rm keV$ per particle. Most of these models have
considered that this energy was injected before the formation of
clusters and groups but, considering the degeneracy between the
redshift of injection and the value of the energy excess, late
injection cannot be ruled out (see Loewenstein 2000).   

It is interesting to note that \emph{not only} the overall correlation
of the cluster population between luminosity and 
temperature (or the number density of clusters at a fixed
luminosity)  are in
disagreement with the 
observations , but also is 
the resulting structure of a
single cluster evolved adiabatically until a relaxed state (see Evrard
1990, Chi\`eze, Teyssier \& Alimi 1998; 
the observed X-ray core radii are at least an order of magnitude larger
than in the simulated cluster). This
implies that preheating should not only affect the 
statistical correlations between clusters taken as a whole population,
but also the internal structure of a particular 
forming cluster. After the
turn-around, the number density of galaxies should be higher in the central
part of a cluster than in the outer parts, producing a spatial gradient in
the quantity of energy injected by galactic feedback, as
well as in the metal
content of the pre-cluster gas. This heating and enrichment can certainly
have dramatic effects on the subsequent evolution of a parcel of gas, and are
up-to-now only crudely approximated by numerical simulations (see
e.g. Metzler \& Evrard, 1994, NFW95). Thus, the relaxed temperature profile
produced by numerical simulations could be significantly altered in a
non-adiabatic model, producing, for example, a temperature gradient. In fact,
recent ASCA spatially resolved cluster spectra (Markevitch et al 1998) and
cluster hydrodynamics simulations (Frenk et al. 1999) seem to confirm
a non-isothermal TP 
in relaxed clusters, even if ROSAT data may not show this gradient (Irwin,
Bregman \& Evrard 1999).

In this paper, I take the point of view that this non-isothermal TP is real
and determine its spatial variation in a steady-state conduction-driven
model. In the absence of a magnetic field, the electronic heat conduction
should transport energy from hot inner gas to colder outer parts, thus
strongly structuring the spatial behaviour of the temperature. In section
\ref{tempprof}, I write down the non-adiabatic energy conservation equation
in this model and solve it for the temperature, before a comparison to X-ray
observations and simulations of clusters. The next section uses the
ROSAT X-ray brightness profile of A496
(Markevitch et al. 1999) to predict the local polytropic
index and the 
hot gas fraction profile, which in turn are compared to ASCA spatially
resolved data. Section \ref{totalmass} compares
the total mass density profile resulting from hydrostatic
equilibrium (hereafter HSE) hypothesis and the analytical TP derived
before with analytic approximations derived from numerical
simulations.  Finally, section \ref{disc} discusses briefly the possible role
of the magnetic field and the consequent inhibition of thermal
conduction, to investigate the amount of time necessary to reach such
a stationnary state.

Throughout the paper, whenever required, a Hubble constant of $H_0 =
50 \, \rm km \, s^{-1} \, Mpc^{-1}$ is used.
 
\section{Conduction-structured temperature profile in clusters}
\label{tempprof}

\subsection{Assumptions}
\label{hypo}
The mean free path of ions and electrons in the ICM are shorter than
the scale length of interest in a cluster (Sarazin, 
1988). Thus, the ICM will be described in the hydrodynamical approximation,
and its evolution governed by the conservation of mass, momentum and
energy. The collisionally ionized plasma is assumed single-phased and an
ideal gas equation state is taken. Since we want to describe a relaxed state,
a steady-state is assumed (i.e. all the terms involving a time derivative
vanish). Moreover, hydrodynamical simulations have shown that the gas
equilibrium is well described by HSE in the exterior dark
matter potential $\phi$ within a radius $R_{500}$ defined by an interior
density which is 500 times the mean density of the universe (Evrard,
Metzler \& Navarro 1996). In fact, for the sake of simplicity, we assume that the
HSE holds until the virial radius of the cluster (defined
as $R_{200}$ with obvious notations). Momentum conservation reduces then to
the hydrostatic equation. All this modeling is done outside of the cooling
flow radius, which is taken to be a fraction $x_0$ of the virial radius. We
thus neglect the radiative cooling of the gas $\Lambda$. Finally, we also
neglect the reheating term $\Gamma$ in the energy conservation equation. As
highlighted in the introduction, this term cannot in general be
neglected. But, since we describe the final state of equilibrium of a
cluster, what we neglect is the \emph{present-day value} of $\Gamma$,
assuming nothing about the value it took before and during the collapse of
the structure.

The thermal conduction flux is assumed to be given by the classical
Spitzer rate (Spitzer, 1965) modified by an efficiency term $f$
($0 < f \leq 1$) to take into account a possible inhibition of the
conduction (see Sec.~\ref{disc}), giving:
\begin{equation}
\label{heatflux}
{\bf q} = f \kappa_0 T^{5/2} \nabla T,
\end{equation}
where the logarithmic dependance of the Coulomb factor with the density has
been ignored and $\kappa_0$ is a constant.   

\subsection{The energy equation for a non-isentropic conductive gas in
hydrostatic equilibrium}
\label{energeq}

Within these asumptions, the mass and energy conservation can be written:

\begin{equation}
\label{masscons}
\nabla . (\rho {\bf v}) = 0
\end{equation}

\begin{equation}
\label{eqconstot}
\nabla . \left[ \rho  {\bf v}  \left( \frac{v^2}{2} + h + \phi
\right) - {\bf q} \right] = 0
\end{equation}
with $\phi$ the total gravitational potential, ${\bf v}$ the bulk
velocity of the gas, $h$ its specific enthalpy and $\rho$ its density.

Assuming HSE means neglecting the spatial part of the
lagangian derivative of the velocity $\left( {\bf v}. \nabla
 \right){\bf v}$ , i.e. neglecting the terms which contain squares
or higher orders of the velocity. Thus, equation (\ref{eqconstot}) can be
simplified to:

\begin{equation}
\label{eqcons}
\nabla . \left[ \rho {\bf v}  \left( h + \phi
\right) - {\bf q}  \right] = 0
\end{equation}

Assuming spherical symmetry, the mass conservation can be integrated to give:

\begin{equation}
\label{massconsinteg}
r^2 \rho v = A_0.
\end{equation}
where $A_0$ is an arbitrary constant and $v$ the radial velocity.
We can integrate the energy conservation as well, and, using equation
(\ref{massconsinteg}), we obtain ($C_0$ being another arbitrary constant):

\begin{equation}
\label{energconsinteg}
A_0 (h + \phi) - r^2 \kappa_0 T^{5/2} \frac{dT}{dr} = C_0
\end{equation}

Since the gas is considered as perfect ($h = 5 kT/2$), equation
(\ref{energconsinteg}) is a differential equation for the temperature, once the
potential $\phi$ is fixed. The exact solution of this equation is beyond the
scope of this paper, and we will restrict us to a special case which has an
analytical solution. Suppose that the equilibrium is static,
i.e. ${\bf v} = {\bf 0}$. If we insert this condition in equation  
(\ref{eqconstot}), we are left with:
\begin{equation}
\label{divfluxnull}
\nabla .(\kappa (T) \nabla T) =  {\bf 0}.
\end{equation}
In other words, the divergence of the heat flux due to conduction
vanishes. This means that, apart from gravitation, there is no heat source or
sink in the intracluster gas nowadays, since conduction is a transport
process. If a temperature gradient is present, heat conduction \emph{freely}
transports energy from the inner hot parts to the outer colder parts. To
compensate the energy loss of the center, an inflow of matter should appear
which contracts the cluster, since the pressure gradient is still fixed by
the hydrostatic condition (This idea is due to R. Teyssier). If this inflow
is subsonic, the contraction will be adiabatic, thus the TP will still be
structured by the local heat conduction and equation (\ref{tsolution}) should
still be valid. The velocity profile induced and the computation of the exact
loss of energy of the center are beyond the scope of this paper, which
intends only to present the model and compare it to observations (see Dos
Santos, 2000, in preparation)

In spherical symmetry, equation (\ref{divfluxnull}) can be
written:
\begin{equation}
\label{staticenerginteg}
r^2 f \kappa_0 T^{5/2} \frac{dT}{dr} = C_0.
\end{equation}
Once we have fixed two integration constants (two temperatures 
at two different radii, say the inner cooling radius and the virial radius),
equation (\ref{staticenerginteg}) can be integrated to give the TP. After some algebra, and
rescaling the radius in units of the virial radius (i.e. $x = r/R_{200}$), we
obtain:
\begin{equation}
\label{tsolution}
\frac{T(x)}{T_{200}} = \left( 1 +
(\eta^{7/2}-1)\frac{x_0}{x_0-1}\frac{x-1}{x} \right)^{2/7},
\end{equation}
where $\eta = T_0/T_{200}$, $T_0$ being the inner temperature and $T_{200}$
the temperature at $r = R_{200}$. This analytic profile will be called
hereafter Steady-State Conduction-Driven temperature profile (SSCD model).

Rephaeli (1977) already constructed a model where the TP was structured by
electronic conduction. However, at this epoch, it was not clear whether the
ICM was mostly primordial or composed by enriched gas ejected by
galaxies. Thus, in his model, he assumed that the galaxies, today, inject gas
in the ICM. This was done by adding a heating term to the heat transfer
equation (eq. ~\ref{divfluxnull}), which is proportional to the galaxy
density profile, assumed to be a King profile and to be proportional to the
total mass density profile. An analytic temperature profile is found, but
depends on the assumed total mass profile. Here, on the contrary, I assume
that $\Gamma(z=0)$ vanishes (see sec. \ref{hypo}), and thus don't need to
assume a total mass profile. The analytic temperature profile has a
determined shape once two boundary conditions are set. On the other hand,
using only X-ray observed quantities like the X-ray surface brightness
profile (which was not yet known at the epoch of Rephaeli's article), we can
obtain the corresponding total mass profile (see section \ref{totalmass}).

Finally, injecting the self-similar analytical form $T(x)/T_{200} = x^\gamma$
into equation \ref{divfluxnull}, it is trivial to show that the only
solutions have indexes $\gamma = -2/7$ or $0$ (the latter one corresponding
to an isothermal cluster). The general solution differs from the self-similar
one only near the virial radius, and reduces to it when $x_0 = \eta^{-2/7}$.

We next compare the SSCD temperature profile obtained above with X-ray
observations and outputs of numerical simulations. 

\subsection{Comparison to observations and simulations}
\label{obssimcomp}

Obtaining the TP of clusters of galaxies from X-ray
spectroscopic and imaging data is not an easy task. 
ROSAT, with both PSPC detectors, was the first satellite to have enough
spatial and spectral resolution to allow crude TPs to be
obtained. Unfortunately, the spectral sensitivity of the detectors was
negligible above $2.4 \; {\rm keV}$, a temperature well below the mean
temperature of rich clusters. Irwin et al. (1999) have searched without
success for temperature gradients in ROSAT data of rich clusters. They did
not derive the TP, but instead worked with hardness ratio profiles. However,
because of its narrow energy band, ROSAT results are surely biased against
the detection of a temperature decrease, especially if calibration
uncertainties are not taken into account (Markevitch \& Vikhlinin,
1997). However, using HEAO 1 A-2 and Einstein data without spatial
resolution (but with different fields of view), Henriksen \& White
(1996) showed the need for a cold component to fit the 
spectral data in four clusters with cooling flows. The emission measure of
this component was so large that it couldn't be explained by the cooling flow
component, and showed that large quantities of cold gas were lying outside
the cooling flow, thus implying a declining TP (if the outer cold gas
is in virial equilibrium in the cluster potential, which can not be
infered from the spectroscopic data alone).

ASCA has much better spectral capabilities than ROSAT but the PSF correction
is problematic. Nevertheless, a number of groups have published TPs
for clusters. In particular, Markevitch et al. (1998, hereafter M98) found
that 19 relaxed clusters ({\it i.e.}  clusters with circular isophotes and
without obvious substructure), when rescaled to their virial
radius and to a flux-weighted mean temperature, had similar TPs within the
error bars. The median TP declines outwards, the temperature decreasing by a
factor of 2 within half the virial radius. It is not yet clear if these
results are reliable because the PSF correction implies that the temperature
measurements are correlated and the systematic effects are poorly known
(see M98). Nevertheless, the next generation of X-ray
satelites (Chandra, XMM, Astro-E) should probe the TP very soon. We will thus
compare the SSCD temperature model to the M98 data.

\begin{figure}
\centerline{\psfig{figure=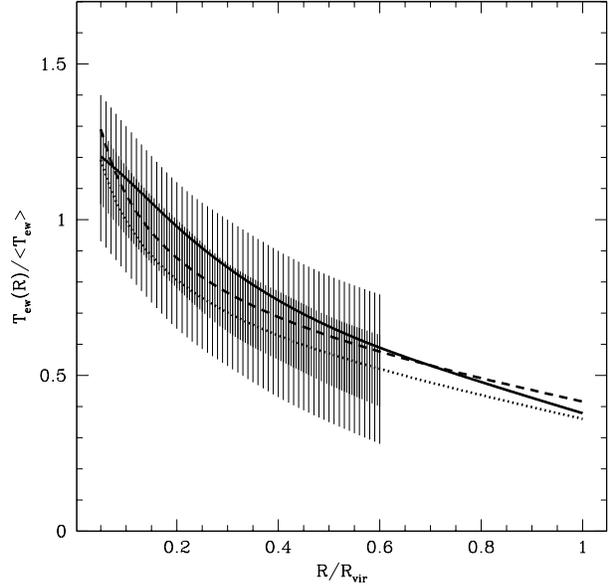,width=8.4truecm}}
\caption{Comparison of the temperature profile with ASCA data of rich
clusters.  The dashed line is obtained with $\eta = 3$ and $x_0 =
0.05$, while the 
dotted line with $\eta=5$ and $x_0 = 0.01$. The light-shaded band shows the
$90 \/ \%$ error boundaries of the M98's composite profile, while the dark
shaded-band corresponds to a less conservative choice, approximating the
scatter of the best-fit profile points.The solid line is the
\emph{mass-weighted} temperature profile obtained from averaging
hydrodynamical simulations of the formation of a cluster (see text).}
\protect\label{tempproffig}
\end{figure}

Remark that the quasi self-similarity of the analytic
solutions naturally explains the fact that all the clusters have the
same temperature profile when rescaled to the virial radius, {\it if the parameter
$\eta$ is roughly constant in rich clusters} (which is not proved
here, but see Dos Santos 2000, in preparation).  To compare properly
the TP with 
M98's observation, we must compute the emission-weighted TP, i.e.:

\begin{equation}
\label{eq:tew}
 T_{ew}(R) = \frac{\int n_g^2 (r) \Lambda [T(r)] T(r) ds}{\int n_g^2 (r)
 \Lambda [T(r)] ds},
\end{equation}
where the integrals are taken along the line-of-sight.  To obtain the gas
density $n_g(r)$, we assume that the surface brightness profile $\Sigma(R)$
is given by a standard $\beta-$model (Cavaliere \& Fusco-Femiano 1976,
1978). This
analytic form gives excellent fits outside of the cooling flow radius (Jones
\& Forman 1984). The two parameters of $\Sigma(R)$, namely the core radius
and $\beta$ are fixed respectively to $R_c = 0.1 R_{200}$ and $\beta =
2/3$. Those values are common in clusters and represent the ``standard''
cluster (Neumann \& Arnaud 1999) used here.  The gas density is obtained by an
Abel inversion integral, with the 
cooling function being pure Brehmsstrahlung ($\Lambda (T) \sim T^{1/2}$), and
then the emission-weighted TP is computed. The results are displayed in
fig. \ref{tempproffig}. The same parameters are used for the computation of
the total mass profile (sec. \ref{totalmass}).
There is a very good agreement with the observations, even if the conductive
profile seems to be less steep in the outer regions of the cluster. Remember
 that there are no data beyond half the virial radius, which means
that the extrapolation is risky. However, interestingly, Markevitch et
al. (1999) note that
the TP for two clusters with high signal-to-noise 
data (A496 and A2199, which are not part of the sample of M98) is more
concave, {\it i.e.} flatter in the outskirts , than their composite
profile. They argue 
that this is a coincidence, but the model presented here explains
naturally the flattening of the TP at large radius (see section
\ref{polytrop}). 

Finally, the SSCD profile is compared to the mean temperature profile
obtained by Frenk et al. (1999) from the average of twelve hydrodynamical
simulations of the same cluster (solid line in figure \ref{tempproffig}). The
virial radius was fixed to $R_{200} = 2.7 \, \rm Mpc$ and the mean
temperature to $7 \times 10^7 \, \rm K$ (somewhat higher than the
mass-weighted mean temperature to mimic emission-weighted mean
temperature). The comparison is not obvious since the simulated TP is
mass-weighted (using an
emission-weighted TP would certainly steep 
the inner simulated profile), the hydrodynamical simulations
are adiabatic, and thus no transport processes such as electronic thermal
conduction are included  and the simulated profile results from an
average of 12 different simulations.  The simulated cluster suffered
its most recent 
strong merger at $z \sim 0.5$. During the phase of violent relaxation
prevailing during a merger, large scale convection should be the main
physical process through which energy is exchanged in the
ICM. But, during the phase of relaxation, once the gravitational
potential suffers no more large fluctuations, the heat
conduction can play a significant local role in the establishment of the
TP. Thus, the simulated temperature profile can be viewed as the initial TP
on which the ETC will act to lead to the SSCD profile. Hence, \emph{perfect
agreement between these two TPs is not expected}. However, at a fixed central
temperature, the gas internal energy in the center is higher in the simulated
TP (keep in mind that mass-weighted and emission-weighted are here
compared, which weakens the argument). The model described in
sec. \ref{tempprof} (transfer by conduction of central energy followed by an
adiabatic contraction) could easily lead to an SSCD profile, without changing
much the central temperature (imposed by the central hydrostatic
condition). 

\subsection{Local polytropic index}
\label{polytrop}

Mixing processes in the ICM,
{\it e.g.} convection, are likely to make the specific entropy constant
within the cluster, and thus lead to an adiabatic structure of the
gas, where pressure and density are simply related by 
\begin{equation}
\label{gasisentrop}
s_0 = c_v \ln \left(\frac{P}{\rho^\gamma} \right),
\end{equation}
$s_0$ being the specific entropy (here considered as constant) and
$c_v$ the specific heat at constant volume. If the gas is perfect,
$\gamma$, the 
ratio of specific heat at constant pressure and volume ($\gamma =
c_p/c_v$) is a constant and its value
(always greater than 1) is fixed by the nature of the gas molecules
: $5/3$ for a monoatomic gas, $7/5$ for
a diatomic gas (see Landau \& lifshitz 1959, p. 315). Even if the
presence of metals in the ICM is spectroscopically important, in view
of the considerable amount of lines they produce, it is
safe to consider that the gas is mainly composed of monoatomic
hydrogen, and thus that $\gamma = 5/3$. This quantity should be kept
constant throughout the ICM, since it depends on the nature of the
plasma itself, and not on its dynamical behaviour\footnote{For gases which can
not be considered as perfect (for example compound of molecules with internal
degrees of freedom excited, like vibrational excitation or ionization,
where the specific heats are not constant), an effective adiabatic
exponent can still be defined formally, but its value is defined by
the variation of the specific internal energy as a function of
temperature and density (in this case, this relation differs from the
one obtained for a perfect gas), which is most conveniently
approximated by a power-law relation $\epsilon \propto T^\alpha
\rho^\beta$. Thus, the value of the effective 
adiabatic exponent will depend on both the exponents $\alpha$ and
$\beta$ of this relation. Surprisingly, its variations are small
compared the the variations of $\alpha$ and $\beta$ for different
gases (see Zel'dovich \& Raizer 1967, pp. 207-209).}. 

\begin{figure}
\centerline{\psfig{figure=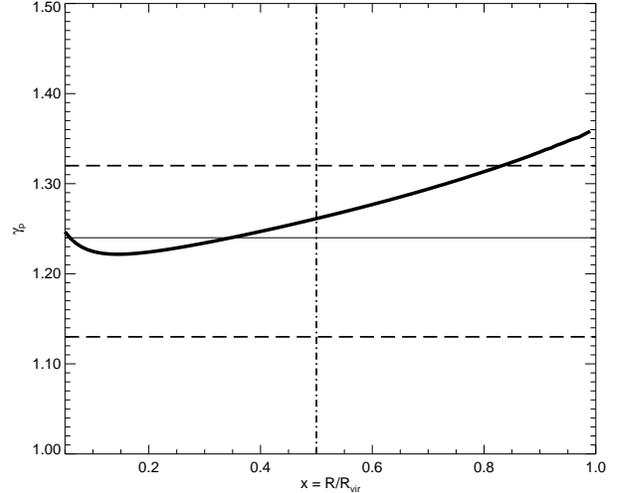,width=8.4truecm}}
\caption{Predicted local polytropic index (bold solid line) compared
to Abell 496 mean polytropic index ($\gamma_p = 1.24^{+0.08}_{-0.11}$,
light solid horizontal line and the two dashed horizontal lines). The
maximum radius where ASCA spectroscopic data were available ($R_{max}
\sim 1.3 \, {\rm Mpc} \equiv 0.5 \times R_{vir}$) is
indicated by the vertical dashed-dotted line. $\eta = 3$ and $x_0
=0.05$ were used for this particular prediction, together with a
$\beta-$model fit to the ROSAT PSPC surface brightness profile (see
text).} 
\protect\label{gammaproffig}
\end{figure}

The first non-isothermal models of the ICM were introduced by Lea
(1975), Gull \& Northover (1975) and Cavaliere \& Fusco-femiano
(1976). They assumed that the ICM was an isentropic perfect gas (with
$\gamma= 5/3$) in equilibrium in a static gravitational
potential. They used 
equation (\ref{gasisentrop}), 
together with the equation of state of the gas, to close the
hydrodynamics equations set and obtain the
TP. The first two-fluid numerical simulations of cluster formation
(Evrard 1990) have shown that the ICM is unlikely to be isentropic,
due to the deepening of the potential well leading to a rising
specific entropy profile with radius. Moreover, the first spatially
resolved 
spectroscopy of clusters have also shown that these adiabatic models
had TP which were too steep, compared to observations (Eyles et
al. 1991, Markevitch et al. 1998). Thus, subsequent non-isothermal
models have often used the following equation as a parameterization of
the temperature profile, after obtention of the density profile via
the HSE equation:
\begin{equation}
\label{gaspolytrop}
P \propto \rho^{\gamma_p}
\end{equation}
In this approach, the polytropic index $\gamma_p$ is a parameter
which is fitted to the spatially resolved spectroscopic data (see {\it
e.g.} Cavaliere et al 1999, figure 5) or used in a two-parameter
models family (Wu et al. 1999, figure 3; Loewenstein 2000). This
parameter is no more related to the microscopic nature of the gas
(this explains why it is called here $\gamma_p$ and not $\gamma$), and
can span a range between 1 (isothermal model) and 5/3 (isentropic gas
and upper limit of the convective stability, Scharzschild
1958)\footnote{In fact, the lower 
limit of $\gamma_p$ is not bounded by some dynamical constraint
(unlike its upper limit), but by the
observational fact that no cluster, outside the cooling flow radius,
has been observed to have an increasing TP with radius.}. Despite the
flexibility of this approach, it is little more than a mathematical
expendiency and its main problem  relies in  that it
links the TP to the gas density profile in an unphysical way: the
density and the 
temperature are forced to track one another in an artificial way,
which can lead to internal inconsistencies when applied to imaging and
spectroscopic X-ray data (Hughes et al., 1988a,b). 
In the present work, on the contrary, the TP is derived from the
resolution of the energy conservation equation, assuming the gas is
perfect and has a constant ratio of the specific heats $\gamma =
5/3$. Then, once the gravitational potential is fixed, this
temperature solution can be inserted in the HSE equation, as in the
papers cited above, to obtain the gas density profile, \emph{without
adding another parameter}. It is thus possible here to predict the
local value of the parameter $\gamma_p$, via the equation
(\ref{gaspolytrop}). Here, $\gamma_p$ is not constant as a function of
radius, but must still be lower than 5/3, in order to ensure
convective stability of the cluster (which would erase any temperature
gradient greater than the adiabatic gradient in some crossing times,
and thus contradict the hypothesis of stationarity). For the sake of
the comparison with TP observations, this method to predict $\gamma_p(r)$
(whose variation will then depend on the gravitational potential
expression) will not be used here. Instead, we  will use the surface
brightness profile fitted to the X-ray data of a cluster to obtain the
gas density profile, assuming that the X-ray photons are emitted via
Brehmstrahlung and the TP is given by equation (\ref{tsolution}).  
The predicted local variations of $\gamma_p$ will be compared to the spatially
resolved spectroscopic data of A496 from Markevitch et al. (1999,
hereafter M99). The
results are qualitatively the same with A2199, the second cluster whose
data are also presented in the paper. The emission-weighted ASCA cooling-flow
corrected temperature is $T_X = 4.7 \, \rm keV$, which gives a virial
radius of $2.67 \, \rm Mpc$, using the relation of
Evrard et al. (1996). The surface brightness profile is taken from
ROSAT PSPC data, and fitted (outside the cooling flow radius) with a
$\beta-$model, giving a core radius of $249 \, \rm kpc \equiv 0.093
\times R_{vir}$ and a slope $\beta = 0.7$. We use these values for the
description of $S_X(R)$ and $(\eta,x_0)=(3,0.05)$ for the TP (using
$\eta=5$ and $x_0=0.01$ instead does not change appreciably
$\gamma_p(r)$). Outside the cooling-flow radius, the TP is described
well by a polytropic fit, with $\gamma_p =
1.24^{+0.08}_{-0.11}$. The figure \ref{gammaproffig} shows the
predicted $\gamma_p$ value (bold solid line), together with the
best-fit value of M99 and its confidence interval (three horizontal
lines, see caption). Despite the fact that no direct fitting of the
TP has been made (the error bars on the data being still large, so we
used the standard parameters of the TP used in section
\ref{obssimcomp}), the agreement with the constant polytropic index
fitted value is impressive within the outer radius of the last ASCA
data bin ($R_{max} \sim 1.3 \, {\rm Mpc} \equiv 0.5 \times
R_{vir}$). Within this radius, the predicted $\gamma_p$ decreases from
1.25 in the center to 1.22 at 400 kpc, then increases again up to 1.36
at the virial radius. The value at $R_{max}$  is 1.26. $\gamma_p$ stays
in the ASCA confidence interval up to $x=0.84$, but is always much
smaller than the adiabatic gradient, which garanties convection
stability within the virial radius.  Note that M99 remark that the
ASCA TP seems more concave than the 
polytropic fit (in fact, the observed TP is flatter than the fit in
the outskirts). The same remark can be made when comparing A496 and
A2199 data with the composite region from 19 clusters derived in
M98. Interestingly, this is exactly what happens in the SSCD
model: within $R_{max}$, the predicted value of $\gamma_p$, although
being very close to the measured value, rises continously from 1.22 to
1.26, which flattens the TP, compared to a constant polytropic
fit. Physically, it is easy to understand why this happens:  the
equation (\ref{divfluxnull}), when integrated over the surface of a
sphere of radius r, stands that the total energy per unit time
crossing the sphere surface is a constant, whatever the radius r (no
source or sinks of energy are present in the ICM). The energy flux $q_r$,
({\it i.e.} the energy crossing a unit surface per unit time) will, on 
the other hand, depend on r as $q_r \propto r^{-2}$ in spherical
symmetry. This flux will be much higher in the center (where the
surface of the sphere is small, but the same amount of integrated
energy crosses it) than in the outskirts. Since the flux is directly
linked to the temperature gradient (equation \ref{heatflux}), the TP
will be much flatter in the outskirts than in the center. Even with
ASCA, the error bars are still too large to allow a discrimination
between the SSCD profile and a polytropic one, but the new generation
of X-ray telescopes (XMM-Newton, Chandra) should be able to settle this
case.   
   
\subsection{hot gas fraction}
\label{hotfrac}

If the gravitational instability picture of the formation of
structures in the universe is crudely right, clusters of galaxies
should be a fair sample of a patch of the 
early universe on a scale of $\sim 10  \, \rm
Mpc$. Thus, the amount of baryonic mass inside a cluster (hot diffuse
gas as well as baryons locked into stars and interstellar medium in
cluster galaxies) divided by the total mass should be also a fair
sample of the baryonic fraction in the universe, since
clusters are the largest structures in equilibrium so far discovered
(if the dynamics of the formation are unable to expell baryonic
material from the deep gravitational potential). This simple idea has
been used as a cosmological test, mainly pointing to the fact that the
measured gas fractions in clusters, if combined  with a standard CDM
universe, were incompatible with the big-bang 
nucleosynthesis results (White et al. 1993). The X-ray emissivity and
the temperature profiles of the ICM, providing a
direct lower limit on the baryonic fraction, are then a very useful
tool to derive 
constraints on the value of the cosmological density parameter (see,
for example, Ettori \& Fabian 1999).       
Most of the recent observational work on X-ray clusters of galaxies
has thus consisted in 
deriving the gas mass profile, in order to obtain a value for the gas
mass fraction (hereafter GMF) at a fixed scaled radius (Evrard 1997;
Arnaud \& Evrard 1999; Mohr et al. 1999; 
Ettori \& Fabian 1999; Vikhlinin, Forman \& Jones 1999) for a large number of
hot clusters (cool clusters or groups of galaxies being more sensitive
to early energy injection and having a shallower potential
well). These studies have shown so far that the 
gas mass profile seems to be similar in all clusters, when rescaled in
proper units (Neumann \& Arnaud 1999; Vikhlinin et al. 1999) and that,
near the virial radius of these hot clusters, the GMF reaches a
constant value, between $14$ and $20 \, \%$ (the value depending
mostly on the method and assumptions used to derive it, {\it e.g.}
isothermal TP), which favours a low value of $\Omega_0$.  
  
\begin{figure}
\centerline{\psfig{figure=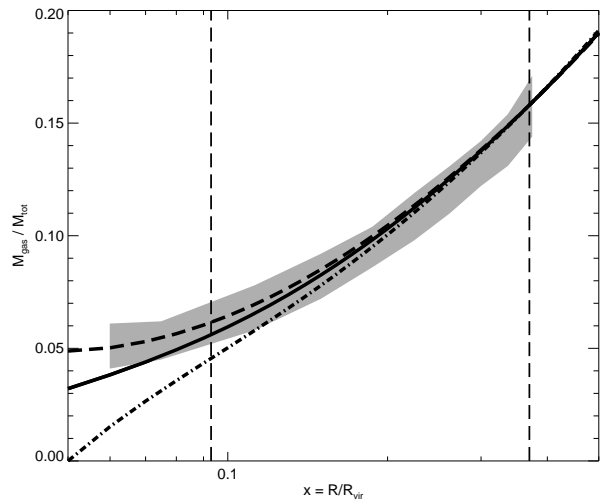,width=8.4truecm}}
\caption{Predicted gas mass fraction profile compared to A496 $90 \,
\%$ confidence interval (shaded region, see M99). See the text for the
meaning of the three curves. The two dashed vertical lines denote the
core radius ($R_c=0.093 \times R_{vir}$) and the radius $R_{1000}$
($\sim 1 \, \rm Mpc$), where the  curves are normalised to an
observed value of $15.8 \, \%$. }
\protect\label{gasfracproffig}
\end{figure}

The derivation of the GMF profile is dependent on the TP, mostly
because the total mass profile depends linearly on the TP. Proceeding
like in the last section, it is then possible to compute the GMF
profile: from the surface brightness profile and the SSCD model, the
gas density profile is computed, which gives the gas mass profile
after spatial integration. The total mass profile (assuming that the
gas mass is negligible, which can be verified {\it a posteriori}) is
derived from the HSE equation. The resulting GMF profile is compared
to the A496 data in figure \ref{gasfracproffig} (the data are
represented by the shaded region, see figure 5 in M99). The predicted
GMF has been normalised at the radius $R_{1000}$ (vertical dashed line
at $x=0.37 \equiv 1 \, \rm Mpc$) at the value given by M99, {\it i.e.}
0.158.  

The SSCD model
predicts a TP which is unbounded in the center, thus the introduction
of a minimum radius $x_0$ (physically identified with the cooling flow
radius). The mass computation is then given by:
\begin{equation}
\label{gasmassform}
M_{gas}(x) = M_{gas}(x < x_0) + m_0\int_{x_0}^{x} {x'}^2 \,
\rho_{gas}(x') \, dx' ,
\end{equation} 
where, $M_{gas}(x)$ is the gas mass profile, $\rho_{gas}(x)$ is the
gas density profile and $m_0$ is a constant. $M_{gas}(x<x_0)$ is the
mass interior to $x_0$, and is also a constant. The model can only
give access to the value of 
the integral in the {\it r.h.s.}. If $M_{gas}(x<x_0)$ is set to zero,
we obtain the lowest curve on figure \ref{gasfracproffig} (dash-dot
line). Here, the gas mass is zero at $x=x_0$, and so is the GMF, which
explains the deviation from the observations. Nevertheless, the
prediction only crosses the boundary of the 90 $\%$ confidence
interval for $x \lesssim 0.1$, and is consistent with the observations
between $x=0.1$ and $x=0.37$ (0.267 and 1 Mpc respectively), but has a
greater slope than them (it is not obvious how much of this effect is
implied by the fact that M99 have used a polytropic TP to compute the
mass profile, but this is probably negligible).

Instead, one can compute the integral in equation \ref{gasmassform}
with a minimum radius much less than $x_0 = 0.05$, say for example 
$x'_0 = x_0/100$ (values below this one don't change much the GMF profile).
Still conserving $M_{gas}(x<x'_0) = 0$, one obtains the solid
curve. The agreement is better, but the discrepancy is still there,
at a lower radius. Finally, one can estimate the constant
$M_{gas}(x<x_0)$, by assuming that the temperature is constant inside
the minimum radius, {\it i.e.} $T(x<x_0) = T(x_0)$. This correction is
applied for the case $x_0=0.05$ only, and gives the dashed
curve. Here, the agreement is perfect, throughout the whole cluster,
the curvature of the predicted GMF being exactly the same as the
observed one. Remark that this curvature is very different than the
one obtained with an isothermal assumption (see M99, figure 5) and
that the only normalisation was on the outer point. This ensures that
the SSCD model (which was not fitted to A496's data, but ajusted ``by eye'' to
the composite profile of M98) describes very well the cluster
spectroscopic results (since the GMF is a derived product of the SSCD
model -- with
more underlying assumptions than only the HSE, which was also used by
M99 -- and the only fitting here was the one of the ROSAT surface
brightness profile performed by M99).      

\subsection{A word on the global and small-scale stability}
\label{stab}

\begin{figure}
\centerline{\psfig{figure=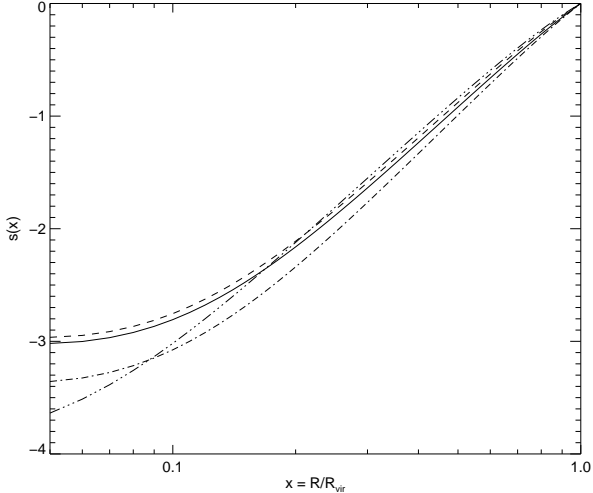,width=8.4truecm}}
\caption{Specific entropy profiles for different sets of parameters
$(\eta,x_0,x_c,\beta)$, where $\eta$ and $x_0$ are the parameters of
the TP (see section \ref{tempprof}) and $x_c$ (in units of the virial
radius) and $\beta$ are the core radius and the third of the
asymptotic slope of the surface brightness profile. The different
lines are solid (3,0.05,0.1,2/3), dashed (5,0.01,0.1,2/3), dot-dashed
(3,0.05,0.093,0.7), and triple-dot-dashed (5,0.01,0.05,0.636). The
last two sets correspond to the surface brightness fits to A496 and
A2199 respectively. The profiles are normalized at $x=1$ and the
specific entropy scale is relative to an unimportant additive
constant.} 
\protect\label{entropprof}
\end{figure}

The question of  the stability of the SSCD
model against general fluctuations is out of the scope of this
paper. I will only briefly comment on this issue. The gas is obviously
stable against large-scale convection instability since, as can be
seen in a particular case in figure \ref{gammaproffig}, the local
polytropic index is, everywhere in cluster, smaller than the adiabatic
value of 5/3. This in turn implies that the specific entropy increases
monotonically with radius (there is a stratification of the gas in the
gravitational potential, according to its specific entropy which
ensures the dynamical stability). This is valid for reasonable values
of the parameters of the TP and of the surface brightness profile. The
figure \ref{entropprof} shows the specific entropy profiles for four
different 
sets of parameters $(\eta,x_0,x_c,\beta)$, namely (3,0.05,0.1,2/3)
our standard cluster depicted as a solid line (section
\ref{obssimcomp}), (5,0.01,0.1,2/3) as a  dashed line,
(3,0.05,0.093,0.7) corresponding to A496, as a dot-dashed line  and
(5,0.01,0.05,0.636) as a triple-dot-dashed line, corresponding to A2199
(see M99). All the entropy profiles increase
with radius, and one can 
see the effect of the core radius $x_c$ and the slope of the surface
brightness profile $\beta$ on the profile, notably the models with the
larger core radii have a constant entropy core.
   
The question of the stability  against small-scale
instabilities is a much harder one.  However, the gas should be
locally stable since, outside the cooling flow, radiation 
cannot enhance density contrasts and thus begin thermal
instability. Moreover,the presence of thermal conduction stabilizes
the plasma against 
small-scale instabilities (Field 1965). This stability is enhanced in the
presence of a weak magnetic field (Balbus 1991), and observations (Bagchi et
al. 1998) as well as simulations (Rocha-Goncalves \& Friaca, 1998)
show that weak magnetic fields should exist in the bulk of the ICM. 

\section{Consequences on the total mass profile in clusters}
\label{totalmass}

The total mass and mass density profile of a cluster are of primary
importance since, if the mass is dominated by dark matter, they can be
directly compared to collisionless numerical simulations. Dubinski \&
Carlberg (1991) found that the relaxed density profile of their simulated
halos were well fitted by a formula derived by Hernquist (1990) from
elliptical galaxy dark matter profiles.  A systematic study of virialized
structures in different cosmological models led Navarro, Frenk \& White
(1996, 1997, hereafter NFW) to propose an analytic expression for the dark
matter density profile, which gives an excellent fit to the
spherically-averaged numerical results, not only in all the cosmologies
explored, but also in a very large range of mass (from galaxies to rich
clusters according to NFW). Further studies (Tormen, Bouchet \& White 1996;
Huss, Jain \& Steinmetz 1999b) extended this result to other cosmologies.
Even if no theoretical basis has been yet established for this dark matter
profile, it could be the result of the violent
relaxation of the dark matter, since collapse with very different initial
conditions give rise to the same profile (Huss, Jain \& Steinmetz 1999a).

However, NFW's inner slope has been recently criticized by
Moore et al. (1998), who find steeper slopes ($r \propto r^{-1.4}$) in their
very high resolution numerical simulations of the formation of clusters, while
Kravtsov et al. (1998) find that a profile with a central core (Burkert
1995) fits better observed and simulated dwarf and low-surface brightness
early-type galaxies. The reasons for these discrepancies are not yet clear.

\begin{figure}
\centerline{\psfig{figure=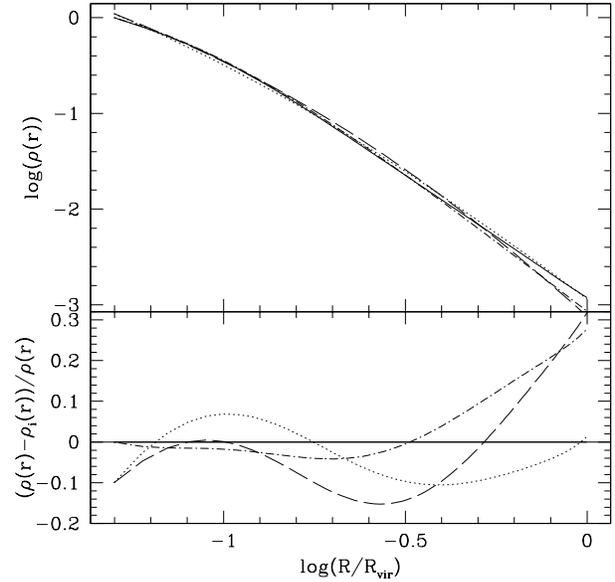,width=8.4truecm}}
\caption{Comparison of the total density profile from the SSCD temperature
profile and the hydrostatic equilibrium (solid line) with NFW profile
(dotted), Hernquist profile (long-dashed) and Burkert profile
(dot-dashed). The lower panel shows the percentage difference between these
three last models and the SSCD profile. For this particular example, $x_0$
was set to $0.05$ and $\eta$ to 3. However, the overall agreement does not
depend much on these values, if reasonable.}
 \protect\label{figmasscomp}
\end{figure}

Assuming a surface brightness profile of the standard cluster and HSE,
the total mass density profile can be computed and the
three analytic functions described above fitted to it. Figure
\ref{figmasscomp} plots the mass density profiles obtained 
and the residuals between the SSCD and analytic profiles.

Although we have made some strong assumptions on deriving the TP, the
agreement is within $16 \; \%$, and even less than $10 \; \%$ in the bulk of
the cluster where temperature information is available. This agreement is
very good, since the residuals to a NFW fit to CDM simulated cluster halos are on
average of $17 \; \%$ and $26 \; \%$ for a Hernquist profile fit (Tormen et
al. 1998). 

From figure \ref{figmasscomp}, it appears that the Burkert profile fits better
the inner regions, while the NFW profile achieves the best fit in the outer
cluster region. \emph{However, this can only be stated with real cluster data, and
depends on the exact parameter pair ($\eta,x_0$) adopted, the overall
$15 \, \%$ agreement being preserved where temperature data are
available}. Moreover, to obtain a good fit, the Burkert profile requires a
very small core radius of the order of $x_0$ ($0.07 \times R_{200}$), which is
equivalent to no core radius. It is not clear if this effect is due to the
core in the analytic X-ray surface brightness profile (since clusters of galaxies
can be fit as well by profiles without core like the Sersic profile; Gerbal,
private communication) or to the steep inner gradient in the SSCD
TP.
\section{Discussion}
\label{disc}

The steady-state conduction model described above predicts the relaxed state of a cluster,
but tells nothing about an important issue, namely the time taken to reach
this steady-state solution. The complete
answer to this important question is beyond the scope of this paper,
and I will only briefly comment this issue.

Obviously, the time taken to reach the relaxed state will depend on one main
factor that governs the inhibition of the heat conduction compared to the
classical Spitzer rate: the inhibition factor $f$. Since the model is
a steady-state one, note that $f$ disappears naturally
from the TP analytic expression when the boundary conditions are
introduced into the general solution. The suppression of
electronic thermal conduction (hereafter ETC) is a longstanding problem in
the theoretical studies of cooling flows (CFs) in clusters of galaxies. The
CF interpretation of the central X-ray properties of a great number of
clusters (a surface brightness excess compared to a $\beta-$model and a
decrease in the central temperature) assumes that gas is removed from the
inflow by thermal instabilities and converted into low-mass stars, in view of
the lack of obvious repositories for the accreted gas within a Hubble
time. This requires a strong suppression of the electronic thermal conduction
(ETC), because of its ability to erase thermal instabilities on timescales
much shorter than the Hubble time. The suppression must be even higher
($f\leq 10^{-4}$) in weakly magnetized multiphased CF models inferred from
observations (Balbus, 1991). There is little doubt that the intracluster
magnetic field plays a primordial role in this inhibition. However, the
traditional point of view that small-scale tangled magnetic fields could
inhibit conduction has been recently severely challenged (Tao 1995, Pistinner
\& Shaviv 1996; see however Tribble 1989) and seems not to be able to provide
enough strong inhibition factors. As already noted by Balbus (1991),
collective plasma effects could provide a viable alternative. In particular,
Pistinner \& Eichler (1998) have recently shown that low-frequency
electromagnetic wave instabilities driven by temperature gradients can
inhibit sufficiently the ETC in CF to reconcile theory and
observations. However, what happens outside the CF is not clear. There, the
cooling time is long enough to ensure that thermal instability will have no
effect and the temperature gradient is much shallower than in the CF. This
last point together with the fact that the magnetic field is expected to
decrease with increasing radius suggests that Pistinner \& Eichler (1998)'s
mechanism is less efficient outside the CF. It is thus worth asking if
a conduction-structured temperature profile can accomodate the X-ray
data and allow then a simple new analytic model of the ICM TP.

Maybe some answers to the above questions will come from X-ray
observations. Chandra recently revealed that in at least two clusters,
A2142 (Markevitch et al., 2000) and A3667 (Vikhlinin, Markevitch \&
Murray, 2000a), dense cool cores are moving with high
velocity through 
the hotter, less dense surrounding gas. A sharp density and
temperature discontinuity (called ``cold front'' by the authors)
separates the 
two phases, while the pressure is continous at the precision level of
the satellite. This phenomenon is very interesting, since we seem to
observe directly the suppression of transport processes in the
ICM. Ettori \& Fabian (2000) have argued that the sharp temperature
discontinuity requires a suppression of heat conduction relative to
the spitzer value by one to three orders of magnitude (depending on
the width of the transition region and on the saturation of the
conduction). The case is 
even stronger for A3667, since the width of the density discontinuity
is smaller than the inferred Coulomb mean free path, showing directly
the suppression of diffusion. In a second paper, Vikhlinin, Markevitch
\& Murray (2000b) argue that the cold front should be
quickly disturbed by Kelvin-Helmholtz instability (while it seems to
be stable in the X-ray images over a $\sim 60$ degrees sector), and
that the stability is ensured by the surface tension of the magnetic
field whose field lines are parallel to the front. In their model, the field lines
are initially frozen into the gas and tangled on some scale on each side of the
front. The stripping of the cool gas stretches the field lines along
the front, which stops the stripping and suppresses
transport processes across the front region. They are able to derive a
value of the magnetic field strength of $\sim 10 \, \rm \mu  G$, an order
of magnitude higher than other estimates of the ICM field (this can be
easily explained by the stretching of the field lines).

This first direct probe of transport processes suppression is very
interesting, but it does not mean  that the same inhibition is
present in all relaxed clusters. In fact, both clusters are in a
dynamically perturbed state: A3667 is classified as a spectacular
ongoing merger in optical (bimodal galaxies distribution and lensing
map), X-rays and radio. A2142 has higly elliptical X-ray isophotes, 
a centroid shift and an asymmetric temperature map. The great
difference in the line-of-sight 
velocities ($\sim 1840 \, \rm km.s^{-1}$) of the two central galaxies 
argues for a dynamically pertubed system, while the presence of a
moderate cooling flow is indicative of the merger being in its late
stages. Thus, it is likely that the phenomenon discovered by Chandra
is much more indicative of what happens during a merger (high magnetic
fields strength, suppression of microscopic transport processes,
convection) than when a
cluster is in a relaxed state for several Gyrs. As has been said in
section \ref{obssimcomp}, I do not expect this model to be valid
during the violent phases of a merger, but much later, when the
cluster had time to relax. Even if a non-negligible percentage of the
cluster population is still dynamically active (particularly the more
massive), their central parts (say, $R \lesssim R_{vir}/2$) should be
described fairly well by the SSCD model. On the other hand, if Chandra
and XMM-Newton, with their improved spatial resolution and
sensitivity, discover the same phenomenon ongoing in a majority of
clusters, even relaxed, the model presented here should not have any
physical basis. Numerical simulations at very high simulation without
thermal conduction indeed show multiple unerased density and temperature
discontinuities for several Gyrs (R. Teyssier, private communication).

\section{Acknowledgements}
This work constitutes part of the PhD thesis of S.D.S. It is a pleasure to
acknowledge useful discussions with Christophe Balland, St\'ephane Colombi,
Florence Durret, William Forman, Christine Jones, Daniel Gerbal, Mark
Henriksen, Barbara Lanzoni, Gast\~ao Lima-neto, Gary Mamon, Maxim Markevitch
and Romain Teyssier. I also thank the referee, V. Eke, for pertinent
remarks which helped improving the clarity of this paper.

\label{lastpage}

\begin{thebibliography}{}

\bibitem[{Arnaud} \& {Evrard}(1999)]{AE99}
{Arnaud}, M., {Evrard}, A.E., 1999, MNRAS 305, 631

\bibitem[{Bagchi} et~al.(1998){Bagchi}, {Pislar}, \& {Lima Neto}]{BPLN98}
{Bagchi}, J., {Pislar}, V., {Lima Neto}, G.B., 1998, MNRAS 296, L23

\bibitem[{Balbus}(1991)]{Balbus91}
{Balbus}, S.A., 1991, ApJ 372, 25

\bibitem[{Bower} et~al.(2000){Bower}, {Benson}, {Baugh}, {Cole}, {Frenk}, \&
  {Lacey}]{BBBCFL2000}
{Bower}, R.G., {Benson}, A.J., {Baugh}, C.M., {Cole}, S., {Frenk}, C.S.,
  {Lacey}, C.G., 2000, MNRAS submitted, astro-ph/0006109

\bibitem[{Bryan} \& {Norman}(1998)]{BN98}
{Bryan}, G.L., {Norman}, M.L., 1998, ApJ 495, 80

\bibitem[{Burkert}(1995)]{B95}
{Burkert}, A., 1995, ApJ 447, L25

\bibitem[{Cavaliere} \& {Fusco-Femiano}(1976)]{CFF76}
{Cavaliere}, A., {Fusco-Femiano}, R., 1976, A\&A 49, 137

\bibitem[{Cavaliere} \& {Fusco-Femiano}(1978)]{CFF78}
{Cavaliere}, A., {Fusco-Femiano}, R., 1978, A\&A 70, 677

\bibitem[{Cavaliere} et~al.(1997){Cavaliere}, {Menci}, \& {Tozzi}]{CMT97}
{Cavaliere}, A., {Menci}, N., {Tozzi}, P., 1997, ApJL 484, L21

\bibitem[{Cavaliere} et~al.(1998){Cavaliere}, {Menci}, \& {Tozzi}]{CMT98}
{Cavaliere}, A., {Menci}, N., {Tozzi}, P., 1998, ApJ 501, 493

\bibitem[{Cavaliere} et~al.(1999){Cavaliere}, {Menci}, \& {Tozzi}]{CMT99}
{Cavaliere}, A., {Menci}, N., {Tozzi}, P., 1999, MNRAS 308, 599

\bibitem[{Chieze} et~al.(1998){Chieze}, {Alimi}, \& {Teyssier}]{CTA98}
{Chieze}, J.P., {Alimi}, J.M., {Teyssier}, R., 1998, ApJ 495, 630

\bibitem[{Dubinski} \& {Carlberg}(1991)]{DC91}
{Dubinski}, J., {Carlberg}, R.G., 1991, ApJ 378, 496

\bibitem[{Ebeling} et~al.(1997){Ebeling}, {Edge}, {Fabian}, {Allen},
  {Crawford}, \& {Boehringer}]{EEFACB97}
{Ebeling}, H., {Edge}, A.C., {Fabian}, A.C., {Allen}, S.W., {Crawford}, C.S.,
  {Boehringer}, H., 1997, ApJ 479, L101

\bibitem[{Eke} et~al.(1998){Eke}, {Navarro}, \& {Frenk}]{ENF98}
{Eke}, V.R., {Navarro}, J.F., {Frenk}, C.S., 1998, ApJ 503, 569

\bibitem[{Ettori} \& {Fabian}(2000)]{EF2000}
{Ettori}, S., {Fabian}, A.C., 2000, MNRAS in press,
  astro-ph/0007397

\bibitem[{Ettori} \& {Fabian}(1999)]{EF99}
{Ettori}, S., {Fabian}, A.C., 1999, MNRAS 305, 834

\bibitem[{Evrard}(1990)]{Ev90}
{Evrard}, A.E., 1990, ApJ 363, 349

\bibitem[{Evrard} \& {Henry}(1991)]{EH91}
{Evrard}, A.E., {Henry}, J.P., 1991, ApJ 383, 95

\bibitem[{Evrard} et~al.(1996){Evrard}, {Metzler}, \& {Navarro}]{EMN96}
{Evrard}, A.E., {Metzler}, C.A., {Navarro}, J.F., 1996, ApJ 469, 494

\bibitem[{Eyles} et~al.(1991){Eyles}, {Watt}, {Bertram}, {Church}, {Ponman},
  {Skinner}, \& {Willmore}]{EWBCPSW91}
{Eyles}, C.J., {Watt}, M.P., {Bertram}, D., {Church}, M.J., {Ponman}, T.J.,
  {Skinner}, G.K., {Willmore}, A.P., 1991, ApJ 376, 23

\bibitem[{Field}(1965)]{Field65}
{Field}, G.B., 1965, ApJ 142, 531

\bibitem[{Frenk} et~al.(1999){Frenk}, {White}, {Bode}, {Bond}, {Bryan}, {Cen},
  {Couchman}, {Evrard}, {Gnedin}, {Jenkins}, {Khokhlov}, {Klypin}, {Navarro},
  {Norman}, {Ostriker}, {Owen}, {Pearce}, {Pen}, {Steinmetz}, {Thomas},
  {Villumsen}, {Wadsley}, {Warren}, {Xu}, \& {Yepes}]{Fetal99}
{Frenk}, C.S., {White}, S.D.M., {Bode}, P., {Bond}, J.R., {Bryan}, G.L., {Cen},
  R., {Couchman}, H.M.P., {Evrard}, A.E., {Gnedin}, N., {Jenkins}, A.,
  {Khokhlov}, A.M., {Klypin}, A., {Navarro}, J.F., {Norman}, M.L., {Ostriker},
  J.P., {Owen}, J.M., {Pearce}, F.R., {Pen}, U.., {Steinmetz}, M., {Thomas},
  P.A., {Villumsen}, J.V., {Wadsley}, J.W., {Warren}, M.S., {Xu}, G., {Yepes},
  G., 1999, ApJ 525, 554

\bibitem[{Gon\c{c}alves} \& {Fria\c{c}a}(1998)]{GF98proc}
{Gon\c{c}alves}, D.R., {Fria\c{c}a}, A.C.S., 1998, in: {Giuricin}, G.,
  {Mezzetti}, M., {Salucci}, P. (eds.), Observational Cosmology: the
  development of galaxy systems, in press, astro-ph/9811264

\bibitem[{Gull} \& {Northover}(1975)]{Gullnorth75}
{Gull}, S.F., {Northover}, K.J.E., 1975, MNRAS 173, 585

\bibitem[{Henriksen} \& {White}(1996)]{HW96}
{Henriksen}, M.J., {White}, R.~E., I., 1996, ApJ 465, 515

\bibitem[{Hernquist}(1990)]{Hern90}
{Hernquist}, L., 1990, ApJ 356, 359

\bibitem[{Hughes} et~al.(1988a){Hughes}, {Yamashita}, {Okumura},
  {Tsunemi}, \& {Matsuoka}]{HYOTM88}
{Hughes}, J.P., {Yamashita}, K., {Okumura}, Y., {Tsunemi}, H., {Matsuoka}, M.,
  1988a, ApJ 327, 615

\bibitem[{Hughes} et~al.(1988b){Hughes}, {Gorenstein}, \&
  {Fabricant}]{HGF88}
{Hughes}, J.P., {Gorenstein}, P., {Fabricant}, D., 1988b, ApJ
  329, 82

\bibitem[{Huss} et~al.(1999a){Huss}, {Jain}, \&
  {Steinmetz}]{HJS99ap}
{Huss}, A., {Jain}, B., {Steinmetz}, M., 1999a, ApJ 517, 64

\bibitem[{Huss} et~al.(1999b){Huss}, {Jain}, \&
  {Steinmetz}]{HJS99mn}
{Huss}, A., {Jain}, B., {Steinmetz}, M., 1999b, MNRAS 308, 1011

\bibitem[{Irwin} et~al.(1999){Irwin}, {Bregman}, \& {Evrard}]{IBE99}
{Irwin}, J.A., {Bregman}, J.N., {Evrard}, A.E., 1999, ApJ 519, 518

\bibitem[{Jones} \& {Forman}(1984)]{JF84}
{Jones}, C., {Forman}, W., 1984, ApJ 276, 38

\bibitem[{Kaiser}(1991)]{K91}
{Kaiser}, N., 1991, ApJ 383, 104

\bibitem[{Kravtsov} et~al.(1998){Kravtsov}, {Klypin}, {Bullock}, \&
  {Primack}]{KKBP98}
{Kravtsov}, A.V., {Klypin}, A.A., {Bullock}, J.S., {Primack}, J.R., 1998, ApJ
  502, 48

\bibitem[{Landau} \& {Lifshitz}(1959)]{LL59}
{Landau}, L.D., {Lifshitz}, E.M., 1959, "Fluid mechanics", Course of
  theoretical physics, Oxford: Pergamon Press, 1959

\bibitem[{Lea}(1975)]{Lea75}
{Lea}, S.M., 1975, Astrophys. Lett. 16, 141

\bibitem[{Loewenstein}(2000)]{Loewenstein2000}
{Loewenstein}, M., 2000, ApJ 532, 17

\bibitem[{Markevitch} et~al.(1998){Markevitch}, {Forman}, {Sarazin}, \&
  {Vikhlinin}]{MFSV98}
{Markevitch}, M., {Forman}, W.R., {Sarazin}, C.L., {Vikhlinin}, A., 1998, ApJ
  503, 77

\bibitem[{Markevitch} \& {Vikhlinin}(1997)]{MV97}
{Markevitch}, M., {Vikhlinin}, A., 1997, ApJ 474, 84

\bibitem[{Markevitch} et~al.(1999){Markevitch}, {Vikhlinin}, {Forman}, \&
  {Sarazin}]{MVFS99}
{Markevitch}, M., {Vikhlinin}, A., {Forman}, W.R., {Sarazin}, C.L., 1999, ApJ
  527, 545

\bibitem[{Metzler} \& {Evrard}(1994)]{ME94}
{Metzler}, C.A., {Evrard}, A.E., 1994, ApJ 437, 564

\bibitem[{Mohr} et~al.(1999){Mohr}, {Mathiesen}, \& {Evrard}]{MME99}
{Mohr}, J.J., {Mathiesen}, B., {Evrard}, A.E., 1999, ApJ 517, 627

\bibitem[{Moore} et~al.(1998){Moore}, {Governato}, {Quinn}, {Stadel}, \&
  {Lake}]{MGQSL98}
{Moore}, B., {Governato}, F., {Quinn}, T., {Stadel}, J., {Lake}, G., 1998,
  ApJ 499, L5

\bibitem[{Navarro} et~al.(1995){Navarro}, {Frenk}, \& {White}]{NFW95}
{Navarro}, J.F., {Frenk}, C.S., {White}, S.D.M., 1995, MNRAS 275, 720

\bibitem[{Navarro} et~al.(1996){Navarro}, {Frenk}, \& {White}]{NFW96}
{Navarro}, J.F., {Frenk}, C.S., {White}, S.D.M., 1996, ApJ 462, 563

\bibitem[{Navarro} et~al.(1997){Navarro}, {Frenk}, \& {White}]{NFW97}
{Navarro}, J.F., {Frenk}, C.S., {White}, S.D.M., 1997, ApJ 490, 493

\bibitem[{Neumann} \& {Arnaud}(1999)]{NA99}
{Neumann}, D.M., {Arnaud}, M., 1999, A\&A 348, 711

\bibitem[{Pistinner} \& {Shaviv}(1996)]{PS96}
{Pistinner}, S., {Shaviv}, G., 1996, ApJ 459, 147

\bibitem[{Pistinner} \& {Eichler}(1998)]{PE98}
{Pistinner}, S.L., {Eichler}, D., 1998, MNRAS 301, 49

\bibitem[{Ponman} et~al.(1999){Ponman}, {Cannon}, \& {Navarro}]{PCN99}
{Ponman}, T.J., {Cannon}, D.B., {Navarro}, J.F., 1999, Nature 397, 135

\bibitem[{Primack} et~al.(1998){Primack}, {Bullock}, {Klypin}, \&
  {Kravtsov}]{PBKK98proc}
{Primack}, J.R., {Bullock}, J.S., {Klypin}, A.A., {Kravtsov}, A.V., 1998, in:
  {Merrit}, {Valuri}, {Sellwood} (eds.), ASP conference series, Galaxy
  Dynamics, astro-ph/9812241

\bibitem[{Rephaeli}(1977)]{R77}
{Rephaeli}, Y., 1977, ApJ 218, 323

\bibitem[{Sarazin}(1988)]{Sarazin88}
{Sarazin}, C.L., 1988, ``X-ray emission from clusters of galaxies'',
Cambridge Astrophysics Series, Cambridge: Cambridge University Press,
1988 

\bibitem[{Schwarzschild}(1958)]{Schwarz58}
{Schwarzschild}, M., 1958, "Structure and evolution of the stars.", Princeton,
  Princeton University Press, 1958.

\bibitem[{Soker} \& {Sarazin}(1990)]{SS90}
{Soker}, N., {Sarazin}, C.L., 1990, ApJ 348, 73

\bibitem[{Spitzer}(1965)]{Spitzer65}
{Spitzer}, L., 1965, ``Physics of fully ionized gases'', Interscience
  Publication, New York

\bibitem[{Tao}(1995)]{Tao95}
{Tao}, L., 1995, MNRAS 275, 965

\bibitem[{Tormen} et~al.(1997){Tormen}, {Bouchet}, \& {White}]{TBW97}
{Tormen}, G., {Bouchet}, F.R., {White}, S.D.M., 1997, MNRAS 286, 865

\bibitem[{Tribble}(1989)]{Tribble89}
{Tribble}, P.C., 1989, MNRAS 238, 1247

\bibitem[{Valageas} \& {Silk}(1999)]{VS99}
{Valageas}, P., {Silk}, J., 1999, A\&A 350, 725

\bibitem[{Vikhlinin} et~al.(1999){Vikhlinin}, {Forman}, \& {Jones}]{VFJ99}
{Vikhlinin}, A., {Forman}, W., {Jones}, C., 1999, ApJ 525, 47

\bibitem[{Vikhlinin} et~al.(2000a){Vikhlinin}, {Markevitch}, \&
  {Murray}]{VMM1}
{Vikhlinin}, A., {Markevitch}, M., {Murray}, S.S., 2000b, ApJ
  submitted, astro-ph/0008496

\bibitem[{Vikhlinin} et~al.(2000b){Vikhlinin}, {Markevitch}, \&
  {Murray}]{VMM2}
{Vikhlinin}, A., {Markevitch}, M., {Murray}, S.S., 2000a, ApJ
  submitted, astro-ph/0008499 (VMM)


\bibitem[{Wu} et~al.(1999){Wu}, {Fabian}, \& {Nulsen}]{WFN99}
{Wu}, K.K.S., {Fabian}, A.C., {Nulsen}, P.E.J., 1999, MNRAS in press,
  astro-ph/9907112

\bibitem[{Zel'Dovich} \& {Raizer}(1967)]{Zeldoraizer67}
{Zel'Dovich}, Y.B., {Raizer}, Y.P., 1967, "Physics of shock waves and
  high-temperature hydrodynamic phenomena", New York: Academic Press,
  1966/1967, edited by Hayes, W.D.; Probstein, Ronald F.


\end{thebibliography}
\end{document}